\documentclass[10pt, conference, compsocconf]{IEEEtran}
\usepackage{graphics}                 
\usepackage{graphicx}
\usepackage[cmex10]{amsmath}
\usepackage{amssymb, amsfonts}
\usepackage{algorithmic}
\usepackage{algorithm}
\usepackage{verbatim}
\usepackage{url}
\usepackage{subfigure}
\usepackage{slashbox}
\usepackage{setspace}
\usepackage{multirow}
\usepackage{array}
\usepackage{hhline}
\usepackage{eso-pic}

\ifCLASSINFOpdf
\else
\fi
\newcolumntype{P}[1]{>{\centering\arraybackslash}p{#1\textwidth}}

\begin{document}
%
\title{Improving Link Prediction in Intermittently Connected Wireless Networks by Considering Link and Proximity Stabilities}


\author{\IEEEauthorblockN{Mohamed-Haykel Zayani, Vincent Gauthier and Djamal Zeghlache}
\IEEEauthorblockA{Lab. CNRS SAMOVAR UMR 5157\\ Institut Mines-Telecom, Telecom SudParis\\
Evry, France\\\{mohamed-haykel.zayani, vincent.gauthier,
djamal.zeghlache\}@telecom-sudparis.eu}
}


%



\maketitle

\begin{abstract}
Several works have outlined the fact that the mobility in
intermittently connected wireless networks is strongly governed by
human behaviors as they are basically human-centered. It has been
shown that the users' moves can be correlated and that the social
ties shared by the users highly impact their mobility patterns and
hence the network structure. Tracking these correlations and
measuring the strength of social ties have led us to propose an
efficient distributed tensor-based link prediction technique. In
fact, we are convinced that the feedback provided by such a
prediction mechanism can enhance communication protocols such as
opportunistic routing protocols. In this paper, we aim to bring out
that measuring the stabilities of the link and the proximity at two
hops can improve the efficiency of the proposed link prediction
technique. To quantify these two parameters, we propose an entropy
estimator in order to measure the two stability aspects over
successive time periods. Then, we join these entropy estimations to
the tensor-based link prediction framework by designing new
prediction metrics. To assess the contribution of these entropy
estimations in the enhancement of tensor-based link prediction
efficiency, we perform prediction on two real traces.
Our simulation results show that by exploiting the information
corresponding to the link stability and/or to the proximity
stability, the performance of the tensor-based link prediction
technique is improved. Moreover, the results attest that our
proposal's ability to outperform other well-known prediction
metrics.

\end{abstract}


%
\IEEEpeerreviewmaketitle

\section{Introduction}
Disruption Tolerant Networks (DTN) paradigm is an emerging wireless
networking application where we have to deal with sparse and
intermittent connectivity. In order to achieve a reasonable packet
delivery rate, we have to rely on opportunistic or mobility-assisted
routing, where messages are forwarded only when two nodes are in
contact. Consequently, the packet delivery rate at the destination
is strongly tied to the network structure during the forwarding
process. Basically, as these networks are human-centered, the
mobility patterns are governed by human behavior. Such a behavior
highly impacts on the structure of the network as shown in
\cite{Chaintreau07,Hossmann2010a}. Moreover, it has been
demonstrated in \cite{Hsu2009a, Thakur2010,Yang2010} that the human
mobility is directed by social intentions that the network users
share at the spatial and temporal levels. When the intentions of
some people are correlated (to be present in the same locations at
the same time), this favors their meeting and thereby the occurrence
of links between them. In this way, mathematical models have been
proposed to characterize the inter-contact time between two people
through statistical analysis \cite{Chaintreau07,Karagiannis2007}.
Thus, it is crucial to better understand how links are created and
to track their properties in order to design efficient communication
protocols.

To analyze the network topology evolution, it is important to rely
on records that describe the status of each link over time during a
tracking period \cite{Braha2006}. Extracting information about
correlations between the willingness of people carrying the nodes is
an important support which provides insights for predicting links.
From this perspective, we have proposed in \cite{Zayani2011} a
tensor-based link prediction technique. Our approach is based on a
spatio-temporal framework that tracks the contacts between nodes.
Hence, tracking the occurrence of links over successive time periods
has enabled us to detect the degree of spatial closeness between the
network users and then quantify their behavior similarity.
Afterwards, this parameter has been used as an indicator to predict
the occurrence of links in the immediate future.

In this paper, we aim to improve the performance of the tensor-based
link prediction technique by refining the measure of behavior
similarity. As we are convinced that link prediction enhances the
performance of communication protocols, we want the feedback
provided by our framework to be the most reliable possible. We find
that the link and the proximity (at two hops) stabilities can also
be considered as parameters to predict future links. Indeed, when
some network users are related by strong social ties, the link
between them tends to be persistent. In other words, when two
network users have high correlated behaviors, their closeness is
expected to lengthen and to remain stable. When nodes have
information about their neighbors at two hops, the stability of such
proximity, joined to the tensor-based link prediction method, also
leads to the performance of more efficient link prediction. We
detail how proximity stability can be beneficial later. Besides,
measuring such stability measures is a key parameter which is
interesting to exploit. To quantify them, we propose a metric
inspired by the Lempel-Ziv estimator \cite{Ziv1977} which converges
to the entropy of a time series. To enhance the prediction
performance of the tensor-based link prediction framework, we join
the stability measures with its outputs by designing new prediction
measures that identify future links as the ones which are relative
to users that have strong (behavior similarity) and stable
(proximity stability) closeness. Afterwards, we assess if our
proposal really contributes to improve the prediction performance of
the tensor-based link prediction method.

The paper is organized as follows: Section II briefly presents the
related work. In Section III, we give an overview of the
tensor-based link prediction technique. Section IV introduces the
measurement of the link and the proximity stabilities, presents the
estimator used to approximate them and details some combinations of
the stability measures with our link prediction technique in order
to provide new prediction metrics. In section V, we present the
simulation scenarios used to evaluate the contribution of these
stability measures, define how to evaluate them and we analyze the
obtained results. Finally, we conclude the paper in Section VI.

\section{Related Work}
Due to the opportunistic aspect of forwarding in intermittently
connected wireless networks and as these networks are basically
human-centered networks, Social Network Analysis (SNA)
\cite{Wasserman1994} has been used in order to provide more
efficient communication protocols. It has been applied to track and
understand the relationships between the network entities and to
extract structural information about the network (network
robustness, topology variance, emerging communities, \ldots).
Several communication protocols such as \cite{Hui2008, Daly2007,
Hossmann2010} have been based on SNA: they have used the centrality
metrics proposed in
\cite{Wasserman1994,Page1999,Hwang2008,Chung1997} and/or have
exploited community detection mechanisms such as
\cite{Wasserman1994,Bollobas1998,Newman2006,Palla2005}. These two
categories of ``social tools'' have been defined as the two major
concepts of SNA in the design of wireless ad-hoc network protocols
by Katsaros et al. \cite{Katsaros2010a}.

Tracking the social ties between the network entities has led to the
design of techniques for link prediction. The link prediction in
social networks has been addressed in the data-mining context
\cite{Acar2009,Dunlavy2011,Wang2007,Liben-Nowell2007} and recently
for the community-based communication networks
\cite{Zayani2011,Wang2011}. These works have highlighted salient
measures that make possible to predict wireless links between
network users. These metrics determine if an occurrence of a link is
likely by quantifying the degree of proximity of two nodes (Katz
measure \cite{Katz1953}, the number of common neighbors, the
Adamic-Adar measure \cite{Adamic2003}, the Jaccard's coefficient
\cite{Jaccard1901,Salton1986}, \ldots) or by computing the
similarity of their mobility patterns (through such metrics as the
spatial cosine similarity, the co-location rate, \ldots). The
efficiency of the Katz measure has been especially emphasized
compared to other prediction metrics in
\cite{Liben-Nowell2007,Wang2011}. Thanks to the SNA, it has been
highlighted that the relationships between individuals has a major
impact on the structure of the network
\cite{Chaintreau07,Hossmann2010a}.

For the human-centered wireless networks, the social closeness
between some people influences their mobility patterns. In
\cite{Song2010}, Song et al. have demonstrated that human mobility
is potentially predictable in 93\% of cases by using the traces of
mobile users of a cellular phone network. This limit has been
investigated through an estimation of the entropy proposed by Ziv
and Lempel \cite{Ziv1977}. The entropy estimation has been applied
on the sequence of visited locations for each mobile phone user
during a tracking period. Song et al. have been motivated by the
findings of \cite{Navet2008} which has highlighted that the entropy
is a very appropriate metric to measure the degree of predictability
of such sequences.

In this paper, we aim to stress that quantifying the degree of
proximity of two nodes joined to the stability of their relationship
enables us to enhance the link prediction performance of
\cite{Zayani2011}. For this objective, we redesign the tensor-based
link technique by taking into consideration the feedback provided by
measuring the link and the proximity at two hops stabilities. Hence,
we describe how to quantify the stability and we propose some
designs for new prediction metrics.

\section{Tensor-Based Link Prediction Framework}
The human mobility patterns highlight correlations in the behavior
of network users. The researches done in
\cite{Chaintreau07,Hsu2009a} have emphasized that the human mobility
depicts a spatio-temporal regularity and the claims advanced in
\cite{Hossmann2010a,Thakur2010} have demonstrated that social ties
characterizing the relationships between users are translated by
correlations between human mobility patterns. From the perspective
that these relationships govern how the network is structured, the
tensor-based link prediction framework aims at identifying and
exploiting these correlations to perform prediction. This is made by
tracking and quantifying the degree of proximity of each pair of
nodes. If two nodes have common intentions at the spatio-temporal
level, they tend to be closer to each other and then nearby. Thus, a
link occurrence between them is likely.

Predicting future links based on their social closeness is a
challenge that is worth an investigation. Indeed, a good link
prediction technique contributes to improving the opportunistic
forwarding of packets and also enhances the delivery rate and/or
decreases latency. Moreover, it helps to avoid situations where
packets overload the queue of the nodes that are unable to forward
these packets towards their final destinations. Motivated by the
enhancement that can provide the prediction to communication
protocols, we propose the tensor-based link prediction framework
which we describe and explain in this section.

\subsection{Notation}
Scalars  are  denoted  by  lowercase letters, e.g., $a$. Vectors are
denoted  by boldface lowercase letters, e.g., $\bf{a}$. Matrices are
denoted   by  boldface  capital  letters,  e.g.,  $\mathbf{A}$.  The
$r^{th}$  column  of a matrix $\mathbf{A}$ is denoted by $\bf{a_r}$.
Higher-order tensors are denoted by bold Euler script letters, e.g.,
$\boldsymbol{\mathcal{T}}$.  The  $n^{th}$ frontal slice of a tensor
$\boldsymbol{\mathcal{T}}$  is  denoted $\mathbf{T_n}$. The $i^{th}$
entry  of  a  vector  $\bf{a}$  is  denoted  by $\bf{a}(i)$, element
$(i,j)$  of  a  matrix $\mathbf{A}$ is denoted by $\mathbf{A}(i,j)$,
and    element    $(i,    j,    k)$    of   a   third-order   tensor
$\boldsymbol{\mathcal{T}}$  is  denoted  by  $\mathbf{T_{i}}(j, k)$.

\subsection{Overview on Tensor-Based Link Prediction Technique}
In order to quantify the degree of spatial closeness of two nodes,
we compute the Katz measure \cite{Katz1953}. It is used in
sociometry and, in the case of wireless networks, expresses the
similarity of the behavior (i.e. mobility patterns) or the degree of
proximity of two nodes. The Katz measure is dependent on the lengths
of paths (one-hop or multi-hop paths) that separate these two nodes.
It is computed from a third-order tensor which records the network
statistics (i.e. occurrence of links between each pair of nodes
during different tracking periods). Hence, a tensor
$\boldsymbol{\mathcal{Z}}$ consists in a set of adjacency matrices
which form successive slices. Each slice corresponds to the contacts
which occurred during a period of time $t$ ($\forall t$, $1 \leq t
\leq T$ where $T$ is the total number of periods). Then, we
determine a collapsed weighted tensor $\mathbf{X}$ which allocates
weights to the links according to the frequency of their occurrence
and their recentness. The use of such a collapsing way is motivated
by the results highlighted in \cite{Acar2009}. Indeed, predicting
links in the period $T$+1 using such a method to aggregate the data
achieved best link prediction performance. Applying the Katz measure
on the collapsed weighted tensor allows us to obtain the matrix of
scores $\mathbf{S}$ which quantifies the behavior similarity or the
degree of proximity for each pair of nodes. An example provided by
Fig. \ref{Zayani0} details how the network structure is tracked and
how the prediction scores are determined for the period $T$+1.

\subsection{Matrix of Scores Computation}
For each node, computing the matrix of scores is performed through
two salient steps. Firstly, the link prediction framework records
the adjacency matrices in a tensor $\boldsymbol{\mathcal{Z}}$ and
determines the collapsed weighted tensor (or matrix) $\mathbf{X}$.
Secondly, it applies the Katz measure on the matrix $\mathbf{X}$ to
obtain the matrix of Katz scores $\mathbf{S}$.

We consider that the data is collected into the tensor
$\boldsymbol{\mathcal{Z}}$. The slice $\mathbf{Z_{p}}(i, j)$
describes the status of a link between a node $i$ and a node $j$
during a time period between $[(p-1) \cdot t,p \cdot t[$ ($p$>0)
where $\mathbf{Z_{p}}(i, j)$ is 1 if the link exists during the time
period $p$ and 0 otherwise. The tensor is formed by a succession of
adjacency matrices $\mathbf{Z_{1}}$ to $\mathbf{Z_{T}}$ where the
subscript letters designate the observed period. To determine the
collapsed weighted tensor, we apply the following expression:

\begin{equation}
    \mathbf{X}(i,j)=\sum_{p=1}^{T} (1-\theta)^{T-p}\ \mathbf{Z_{p}}(i,j)
    \label{eq1}
\end{equation}
where the matrix $\mathbf{X}$ is the collapsed weighted tensor of
$\boldsymbol{\mathcal{Z}}$, and $\theta$ is a parameter used to
adjust the weight of recentness and its value is between 0 and 1.

\begin{figure}[!tb]
    \centering
    \includegraphics[width=0.5\textwidth]{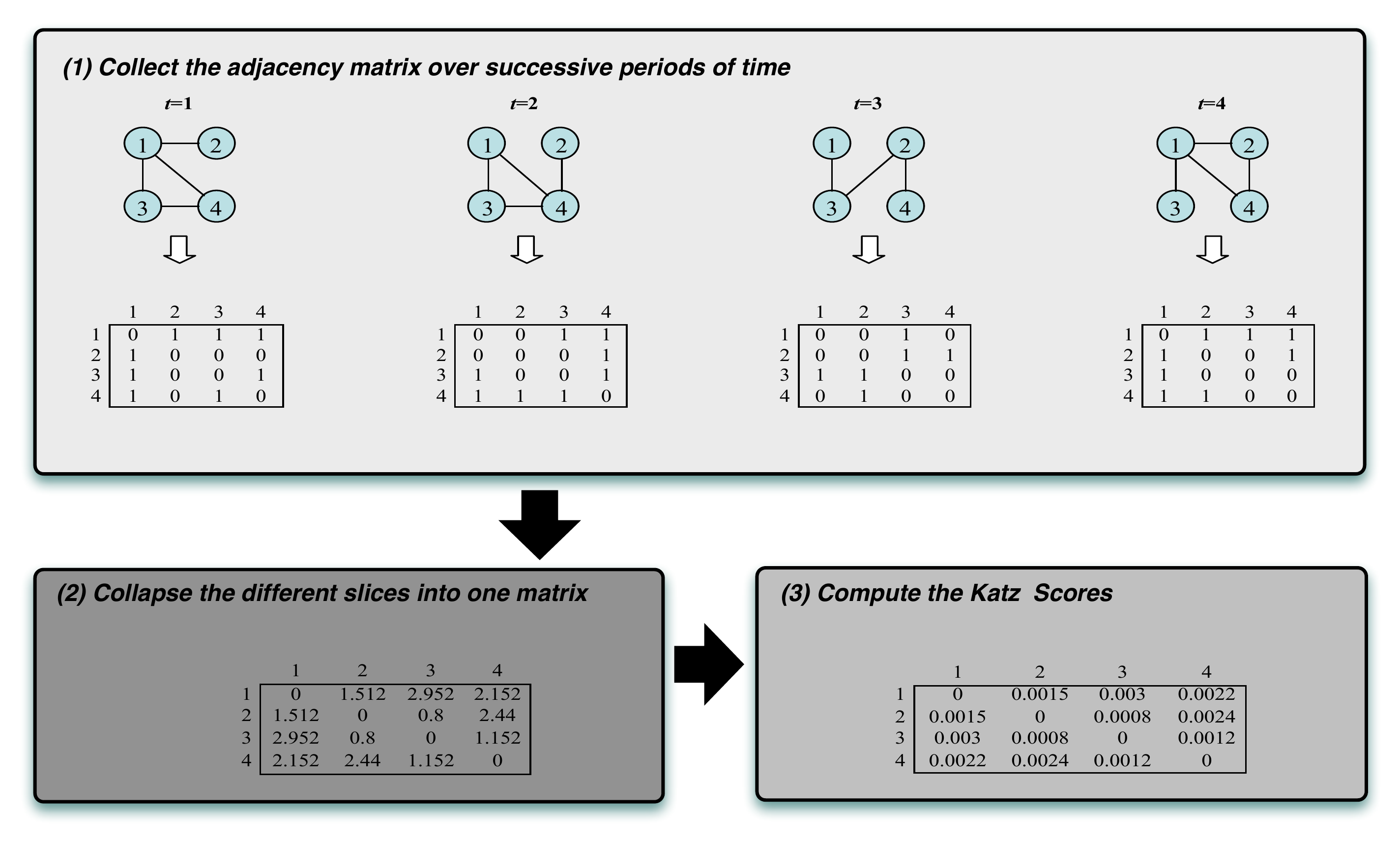}
    \caption{Example of the matrix $\mathbf{S}$ computation}
    \label{Zayani0}
\end{figure}

The matrix $\mathbf{X}$ can be seen as a generalized adjacency
matrix with weighted links. A node can determine the behavior
similarity with the other nodes by computing the Katz measure. This
similarity between two nodes $i$ and $j$ is determined by the length
of paths that connect $i$ to $j$ and is impacted by the weight of
each path length.

Then, Eq.(\ref{eq3}) is used to compute matrix of Katz scores
$\mathbf{S}$ as following:

\begin{equation}
    \mathbf{S}=\sum_{\ell=1}^{+\infty} \beta^{\ell} \cdot \mathbf{X}^{\ell}=(\mathbf{I}-\beta \cdot \mathbf{X})^{-1}-\mathbf{I}
    \label{eq3}
\end{equation}
where $\beta$ is a user defined parameter which is strictly positive
and $\beta^{\ell}$ is the weight of a $\ell$ hops path length. It is
clear that the longer the path is, the lower its weight gets. The
matrix $\mathbf{I}$ is the identity matrix and $\mathbf{X}$ is the
obtained collapsed weighted tensor.

We depict as previously mentioned in Fig. \ref{Zayani0} an example
which details the two major steps described before. We take into
consideration a network consisting of 4 nodes and having a dynamic
topology over 4 time periods and we highlight how similarity scores
are obtained ($\theta$ and $\beta$ are respectively set to 0.2 and
0.001 for the example and later for the simulations). In this
example, we assume that all nodes have the full knowledge of the
network structure.

After the investigation into the value to be chosen for the
parameter $\beta$, we found that the convergence of the Katz measure
is closely tied to the spectral radius of the adjacency matrix, as
mentioned in \cite{Franceschet2011}. In fact, $\beta$ must be
greatly inferior to the reverse of the latter value. In our case, we
use the collapsed weighted tensor which can be considered as an
adjacency matrix with weights of links. After testing the
performance of our method through different values of $\theta$ and
$\beta$, we found that setting $\theta$ to 0.2 and $\beta$ to 0.001
enables us to reach one of the highest prediction performances. For
more details, we refer the reader to \cite{Zayani2011} in which we
looked at the performance of our framework with different values
$\theta$ and $\beta$.

\subsection{Matrix of Scores Interpretation}
The social ties between each pair of nodes are quantified by the
similarity measure given by $\mathbf{S}(i,j)$. When two nodes share
an important score, this means that the paths that connect them tend
to be short. Two nodes separated by few hops rimes with a
geographical closeness which underlines a correlation between their
behavior and emphasizes a spatial proximity. Therefore, the link
occurrence between these two nodes is strongly plausible. Otherwise,
if the similarity score is low or null, the two corresponding nodes
share occasional or even no correlation in their behavior.

\section{How To Quantify Link and Proximity Stabilities and How To Use Them?}

As detailed previously, to predict links in intermittently connected
wireless networks, we have used a tensor-based framework to record
network structure for a number of periods and have measured Katz
scores for each pair of nodes to quantify their behavior similarity.
In \cite{Zayani2011}, we have shown that the tensor-based link
prediction framework achieves similar performance either when done
in a centralized way (i.e. assuming that a central unit records the
network statistics and performs prediction) or in a distributed way
(i.e. the nodes predict links relying on local information).
Obviously, we are interested in the distributed application of the
prediction framework as we are concerned by the intermittently
connected wireless networks and the effectiveness of our proposal in
this case is strongly prominent. Nonetheless, we strongly conceive
that we can further improve the prediction efficiency. We think that
the prediction becomes more precise if we take into consideration
other salient characteristics of ties between nodes. Indeed, we
guess that focusing on the stabilities of links and of proximity at
two hops are worth investigating and the predictability of human
mobility made in \cite{Song2010} has further motivated us to
continue in this direction.

\subsection{How Can Link and Proximity Stabilities Improve Link
Prediction?}

Regarding link stability, when two nodes have strong social ties, a
link occurrence between them is likely. If a link between them
occurs, it is expected to be persevering and then stable. Indeed,
they are expected to be close for a fairly substantial time as they
share similar behavior. On the other hand, when the proximity at two
hops between two nodes is stable, it can be interpreted in two
separate manners. When the stability is expressed by the absence of
proximity at two hops, this means that the two nodes are either tied
by a link or separated by more than two hops. In this case, the
outputs of the tensor-based link prediction technique can be used to
identify if the corresponding nodes are directly tied. When this
stability is expressed by two nodes constantly separated by two
hops, the information provided by the tensor-based link prediction
technique is able to attest that a link occurrence is unlikely.
Moreover, the corresponding Katz score is expected to be lower than
those of pair of nodes having a link between them. If the proximity
is unstable, it is obvious to conclude that the link occurrence is
unlikely. Indeed, even if there is a link between the two nodes, its
occurrence is intermittent and then the link is unstable. The
information provided by the Katz measure gives indications about the
real state of the link.

Hence, measuring such stability parameters is paramount to enhance
the performance of the tensor-based link prediction framework.
Therefore, we should refine the tensor-based link prediction
framework to make it sensitive to these parameters.

\subsection{Quantifying Link and Proximity Stabilities by Means of Time Series Entropy Estimation}

In order to measure these stabilities, we opt for the entropy
metric. It is well-known that this measure has been widely used to
quantify stability or uncertainty in several domains such as
thermodynamics and information theory
\cite{Boltzmann1877,Shannon1948}. For our approach, we have been
interested in the entropy estimator used for the Lempel-Ziv data
compression \cite{Ziv1977,Kontoyiannis1998}. This tool enables us to
estimate the entropy of a time series. It has been applied by Song
et al. \cite{Song2010} on the sequences of locations visited by
several cellular network clients to demonstrate that human mobility
is highly predictable. For a record of $n$ steps, the entropy is
estimated by:

\begin{equation}
    S^{est}=\left ( \frac{1}{n}\sum_{i}\Lambda_{i}  \right )^{-1} \ln n
    \label{eq_lempel_ziv}
\end{equation}
where $\Lambda_{i}$ is the length of the shortest substring starting
at position $i$ which does not previously appear between position 1
and $i-1$.

Before exploiting this estimator in link prediction, we have to
define the time sequences in order to measure the stability of each
link. Instead of Song's et al. approach which consists in filling in
the user's location labels for each step, we rely on statistics of
the link state between each pair of nodes. To construct this
sequence, each node has to record the state of a link with every
detected neighbor and at each period. Tracking the status of each
link through a third order tensor accurately achieves this
requirement and for each link we obtain a sequence of zeros and/or
ones. Therefore, we propose the entropy estimator $E^{l}_{T}(i,j)$
which quantifies the stability of the link between the nodes $i$ and
$j$ over $T$ periods. This is given by:

\begin{equation}
    E^{l}_{T}(i,j)=\left ( \frac{1}{n}\sum_{t=1}^{T}\Lambda_{t}(\mathbf{Z_{t}}(i,j))  \right )^{-1} \ln
    n
    \label{eq_lempel_ziv_2}
\end{equation}
where $\Lambda_{t}(\mathbf{Z_{t}}(i,j))$ is the length of the
shortest substring (consisting of a sequence of zeros and/or ones)
starting at position $t$ which does not previously appear between
position 1 and $t-1$. The parameter $n$ corresponds to the number of
substrings which are identified. 

In the same way, we define $E^{p}_{T}(i,j)$ as estimator of the
entropy which quantifies the stability of the proximity at two hops
of the pair of nodes $(i,j)$. It is computed as in Eq.
(\ref{eq_lempel_ziv_2}) but by substituting, for the value
$\mathbf{Z_{t}}(i,j)$, the state of proximity at two hops at period
$t$ (1 if $i$ and $j$ are separated by 2 hops at the period $t$ and
0 otherwise).

The Lempel-Ziv entropy estimator identifies at each step the
shortest sequence which is not detected before. Therefore, we are
tracking the length of the substrings step by step. If the shortest
substrings become quickly too long, this means that there is
redundancy. Indeed, the first chains added in the set of shortest
substrings are repetitive in the whole sequence and it is necessary
to concatenate new strings to make new shortest chains. Then,
redundancy matches with stability as long chains decrease the value
of the estimator. Otherwise, if the shortest substrings too often
take the smallest possible length, this means that there vastly
different combination of zeros and ones. This remark suggests that
the ties highlight randomness and variation rather than regularity.
So, the status of a tie is more unstable and the entropy estimation
gets higher (due to short length of the substrings).


In the following subsection, we present new prediction scores. They
are determined by joining the entropy estimations to the
tensor-based link prediction framework.

\subsection{Joining the Entropy Estimations to the Tensor-Based Link
Prediction Framework}

We have shown that the measures that we advance can be used to
quantify the stability of a link or the proximity at two hops.
Nevertheless, these measures, as mentioned previously, are unable to
determine if the stability is quantified for an occurring tie
(series of ones in the tensor) or for a tie that is occasionally or
never created (series of zeros in the tensor). Then it is important
to combine them to a metric that expresses the lifetime or the
perseverance of a link. From this perspective, we propose the
combination with the Katz measure and/or the weight provided by the
collapsed weighted tensor. Therefore, after evaluating several ways
to combine the entropy estimations with the tensor-based link
prediction framework, we propose four different techniques to join
them. Our aim is to demonstrate that measuring the stability and
exploiting it are really beneficial to improve the prediction
performance. In this work, we want to prove that considering the
stability of ties between nodes improves the link prediction of our
framework. We are not seeking the design that ensures the best
prediction efficiency. This investigation will be the aim of a
future work. In the following, we detail how we design the
combinations in order to predict better a link occurrence between a
node $i$ and a node $j$ through new metrics.

\begin{table}[t]
\renewcommand{\arraystretch}{1}
\caption{Table of confusion of a binary prediction technique}
\label{table_confusion} \centering \scalebox{0.65}{
\begin{tabular}{|c|P{0.2}|P{0.2}|}
\hline \backslashbox{Prediction outcome} {Actual
value}& Positive & Negative\\
\hline
Positive & True Positive ($TP$) & False Positive ($FP$)\\
\hline
Negative & False Negative ($FN$) & True Negative ($TN$)\\
\hline
\end{tabular}
}
\end{table}

\begin{itemize}
\item \textbf{Combining the collapsed weighted tensor value
$\mathbf{X}(i,j)$ with the link entropy estimation $E^{l}_T(i,j)$
(XE scores):} We suggest to join the entropy estimation
$E^{l}_{T}(i,j)$ (as expressed in Eq. (\ref{eq_lempel_ziv_2}) and
where $(i,j)$ is a pair of nodes) to the weight collected by the
matrix $\mathbf{X}(i,j)$ (Eq. (\ref{eq1})) as we are seeking the
most stable occurring links. In other words, we want to identify the
links that have high weight and in the same time low estimation of
the entropy. We use normalized values for $E^{l}_{T}(i,j)$ and
$X(i,j)$.

\item \textbf{Combining the Katz measure $\mathbf{S}(i,j)$ with
the link entropy estimation $E^{l}_{T}(i,j)$(SE scores):} We proceed
as for the previous suggestion but we join the entropy estimation
with the behavior similarity metric $\mathbf{S}(i,j)$ (Eq.
(\ref{eq3})). We aim to check and assess that joining sociometric
and stability measures can be helpful to make more precise
predictions. As in the previous combination, we use normalized
values for both parameters.

\item \textbf{Computing the Katz measure $\mathbf{S}(i,j)$
from the combination of the collapsed weighted tensor value
$\mathbf{X}(i,j)$ with the link entropy estimation $E^{l}_{T}(i,j)$
(XES scores):} We propose to apply the tensor-based link prediction
technique but combining the collapsed weighted tensor $\mathbf{X}$
with the link entropy estimation. Indeed, we combine each weight
$\mathbf{X}(i,j)$ (Eq. (\ref{eq1})) with the measure
$E^{l}_{T}(i,j)$ (Eq. (\ref{eq_lempel_ziv_2})) and we apply the Katz
formulation on the described combination in order to obtain a new
matrix of scores. Also, we use normalized values for the entropy
estimation and the collapsed weighted tensor values.

\item \textbf{Computing the Katz measure $\mathbf{S}(i,j)$
with the definition of the new collapsed weighted tensor value
$\mathbf{X_{new}}(i,j)$ (XNS scores):} We also propose to apply the
tensor-based link prediction technique but with a new collapsed
weighted tensor denoted $\mathbf{X_{new}}$. To determine it, we
compute a coefficient, at each period $p$ and for each pair of nodes
$(i,j)$, that combines the occurrence weight obtained from
$\mathbf{X}$ (i. e. lifetime and recentness) with the link stability
and/or the proximity stability. Afterwards, we apply the Katz
formulation on the new collapsed weighted tensor $\mathbf{X_{new}}$
in order to obtain a new matrix of scores. The matrix
$\mathbf{X_{new}}$ can be expressed in different ways which are
detailed in the following.

\end{itemize}

Among the scores involving the entropy estimation that we have
tested, we hold from each category the metric/metrics which
highlights/highlight regularity and efficiency for all simulation
scenarios. Then, for $T$ tracking periods, we define the following
matrices of scores:

\begin{eqnarray}
XE\_Score & = & ([1]_{N \times N}-\mathbf{X})\bullet\times E^{l}_{T}
\label{eq_XE}
\\
SE\_Score & = & ([1]_{N \times N}-\mathbf{S})\bullet\times E^{l}_{T}
\label{eq_SE}
\\
XES\_Score & = & (\mathbf{I}-\beta \cdot [E^{l}_{T}\bullet\times\\
\nonumber & &([1]_{N \times N}-\mathbf{X})])^{-1}-\mathbf{I}
\label{eq_XES}
\\
XNS\_Score & = & (\mathbf{I}-\beta \cdot
\mathbf{X_{new}})^{-1}-\mathbf{I} \label{eq_XNS}
\end{eqnarray}
where $N$ is the number of nodes involved in the statistics and, for
two nodes $i$ and $j$, $\mathbf{X_{new}}(i,j)$ is given by:
\begin{eqnarray}
\mathbf{X_{new}}(i,j)&=&\sum_{t=1}^{T} (1-\theta)^{2(T-t)}\ \cdot
(\mathbf{Z_{t}}(i,j) \cdot \\ \nonumber & &[\max_{t}(E_t)-E_t(i,j)])
\label{xnew_1}
\end{eqnarray}
The parameter $E_t(i,j)$ and $\max_{t}(E_t)$ respectively correspond
to the current entropy estimation (whether for link stability or
proximity stability) and the maximum entropy value that we can
obtain for $t$ periods.

We derive three variants of the $XNS\_Score$. When $E^{l}_t(i,j)$ is
used to compute $\mathbf{X_{new}}(i,j)$, we define the measure
$XNS1\_Score$. If $E^{p}_t(i,j)$ is chosen, we express the metric
$XNS2\_Score$. In addition, we consider the case in which we compute
$\mathbf{X_{new}}(i,j)$ using the link and the proximity
stabilities. In this case, we propose the measure $XNS3\_Score$
where $\mathbf{X_{new}}(i,j)$ is given by:
\begin{eqnarray}
\mathbf{X_{new}}(i,j)&=&\sum_{t=1}^{T} (1-\theta)^{3(T-t)}\ \cdot
(\mathbf{Z_{t}}(i,j) \cdot \nonumber \\& &
[\max_{t}(E_t)-E^{l}_t(i,j)] \cdot \nonumber \\&
&[\max_{t}(E_t)-E^{p}_t(i,j)]) \label{xnew_2}
\end{eqnarray}

In the following section, we detail the different scenarios to
evaluate all these metrics and we assess their ability to achieve
better performance in link prediction.

\section{Simulations Scenarios and Performance Evaluation}
To evaluate the efficiency of the tensor-based link prediction
method joined to the entropy estimations, we consider two real
traces. We present them in the following subsection. Then, we
analyze the results obtained and we assess the efficiency of our
contribution.

\subsection{Simulation Traces}
We consider two real traces to evaluate the contribution of the
entropy estimations. We exploit them to construct the tensor by
generating adjacency matrices with different time periods lengths.
We detail, in the following, the traces used for the evaluation:
\begin{itemize}
\item \textbf{First Trace: MIT Campus trace:}
We take the trace of 07/23/02 \cite{Balazinska2003} and consider the
events between 8 a.m. and 4 p.m. (8 tracking hours) to build up the
tensors. The number of nodes is 646 and the number of locations
(i.e. access points) is 174.
\item \textbf{Second Trace: Dartmouth Campus trace:}
We choose the trace of 01/05/06 \cite{Dartmouth} and construct the
tensor slices relying on SYSLOG traces between 8 a.m. and 4 p.m.
also. The number of nodes is 1018 and the number of locations (i.e.
access points) is 128.

\end{itemize}
For each scenario (a fixed tracking period length), we track the
occurrences of contacts during $T$ periods. We also consider the
adjacency matrix corresponding to the period $T$+1 as a benchmark to
evaluate the effectiveness of our proposal. We construct tensors for
the following period lengths: 5, 10, 30 and 60 minutes. That is to
say that we record the network statistics for respectively a number
of periods $T$ equal to 96, 48, 16 and 8 slices (for the case where
$t$=5 minutes, it is necessary to have 96 periods to cover 8 hours,
48 periods are needed to do the same when $t$=10 minutes, \ldots).
We consider the distributed case for the computation of scores. We
assume that each node has the knowledge about its 1-hop and 2-hop
neighbors to compute the Katz measure for the tensor-based link
prediction technique and the proposed measures of stability.

\subsection{Simulation Results and Performance Analysis}
To asses if the proposed metrics enhance the prediction performance
of the tensor-based link prediction framework, we consider these
evaluation measures:
\begin{itemize}
\item \textbf{Top Scores Ratio at the period $T$+1 (TSR):} we compute the number of occurring links
in the period $T$+1 (the first period coming after the record of the
network statistics). We call the number of existing links $L$. Then,
we extract the links having the $L$ highest scores found after
applying the prediction technique and determine the percentage of
existing links in both sets.
\item \textbf{Accuracy (ACC):} this measure is defined in \cite{FAWCETT2006} as
the ratio of correct prediction (true positive and true negative
predictions) over all predictions (true positive, true negative,
false positive and false negative predictions). In other words, it
is computed by the ratio $\frac{TP+TN}{TP+FP+TN+FN}$ (see Table
\ref{table_confusion}). We identify for each scenario the maximum
value of the accuracy which indicates the degree of precision that
can reach each prediction metric.
\item \textbf{F-Measure or balanced F1 score:} 
the F-measure \cite{vanRijsbergen1979} is the harmonic mean of
precision\footnote{represents to the proportion of links with
positive prediction (occurring in the future) which are correctly
identified \cite{FAWCETT2006}. Based on Table \ref{table_confusion},
the precision is equal to $\frac{TP}{TP+FP}$. This value is
determined according to the deduced accuracy value.} and
recall\footnote{quantifies the ratio of correctly identified links
over the occurring links in the future \cite{FAWCETT2006}. Referring
to Table \ref{table_confusion}, the recall is defined by the
expression $\frac{TP}{TP+FN}$. This value is also computed according
to the retained accuracy value.}. The F-Measure is expressed by
$2.\frac{precision.recall}{precision+recall}$. The higher the
F-Measure is, the better the tradeoff of precision and recall gets
and the more efficient the prediction metric is.

\end{itemize}

\begin{figure}[!t]
  \centering
  \subfigure[Top Scores Ratio in $T$+1]{
    \label{fig5}
    \includegraphics[width=0.3\textwidth,angle=270]{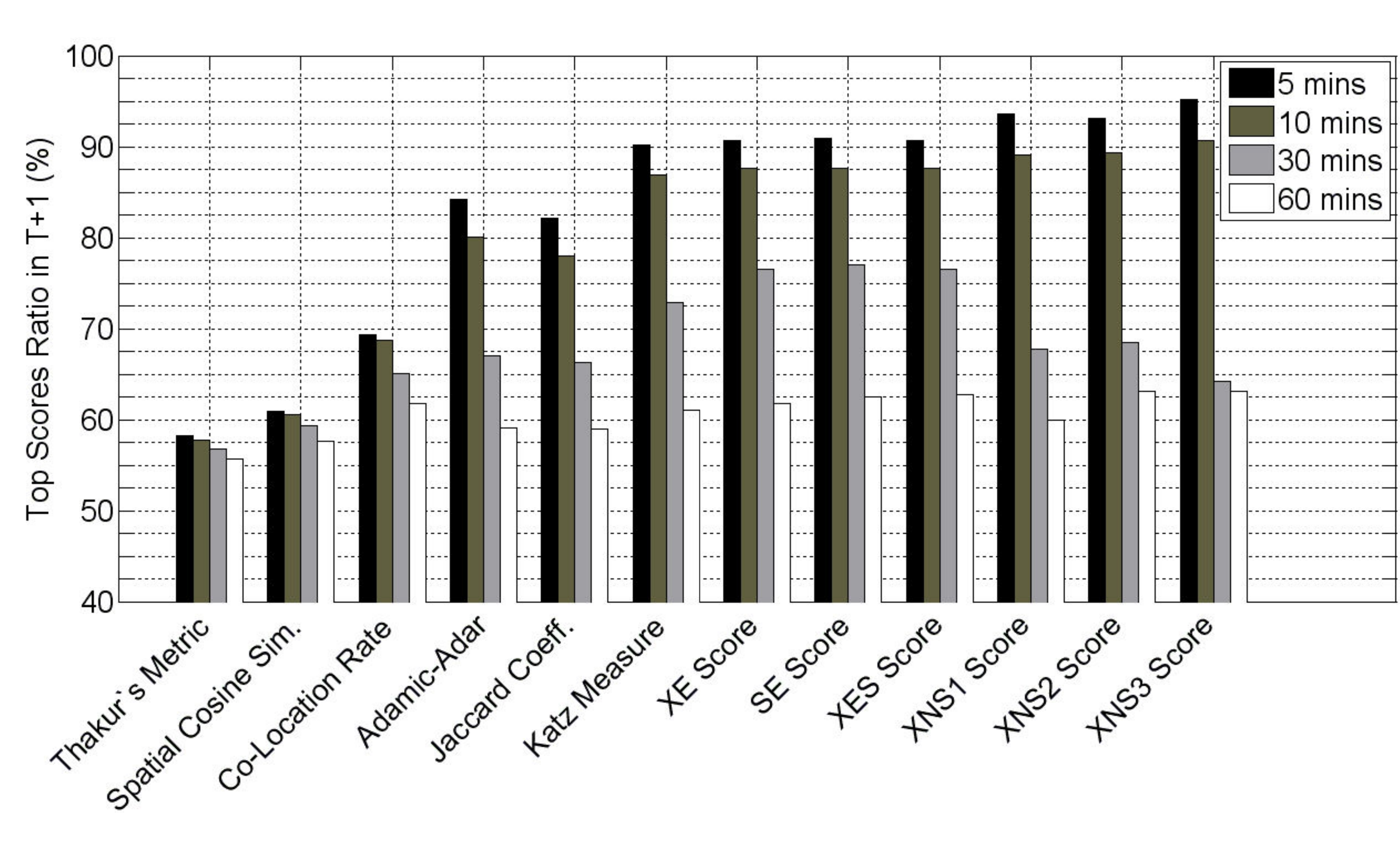}
  }\\
  \subfigure[Accuracy (percentage exceeding 99\%)]{
    \label{fig6}
    \includegraphics[width=0.3\textwidth,angle=270]{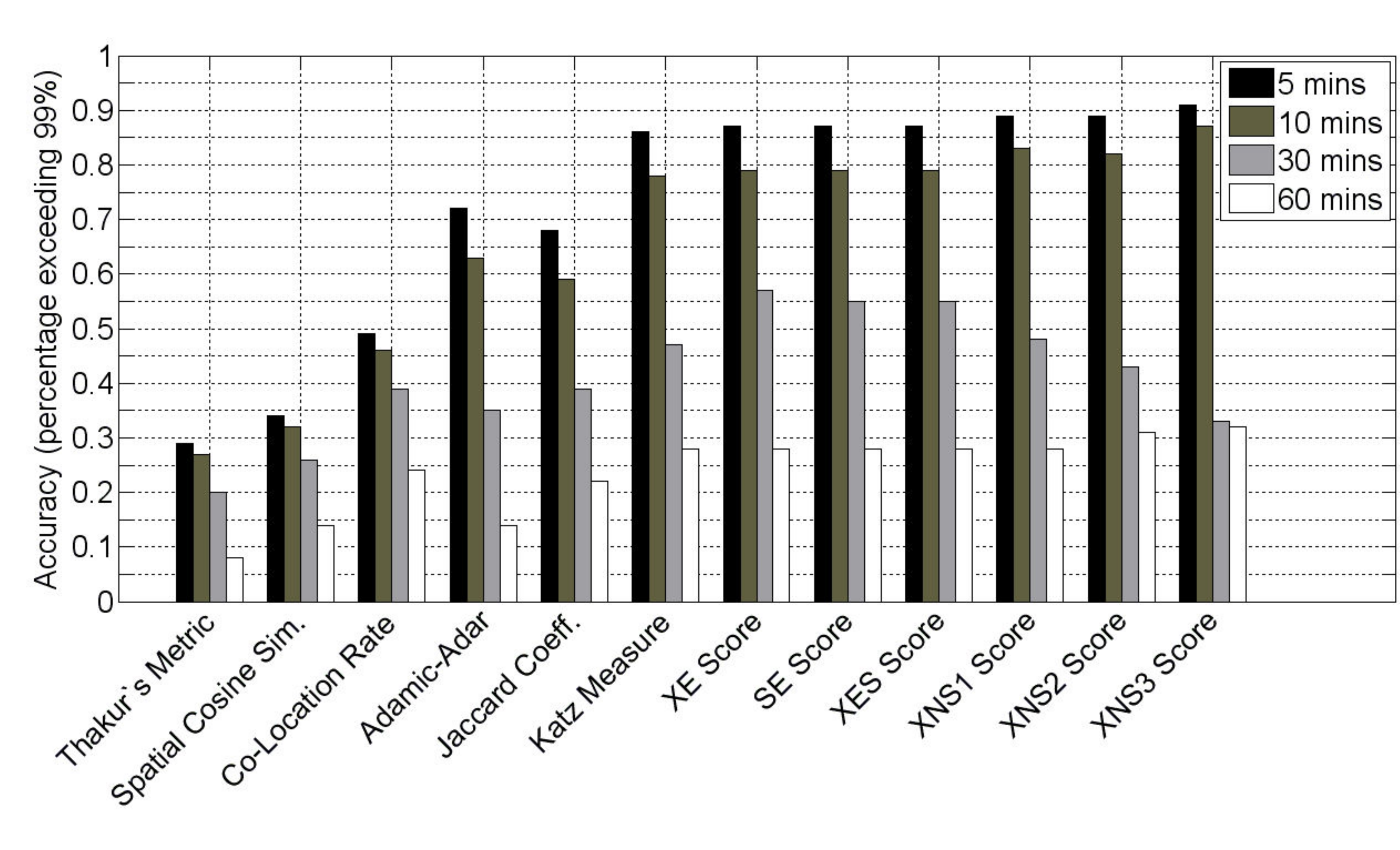}
  }\\
 \subfigure[F-Measure]{
    \label{fig7}
    \includegraphics[width=0.3\textwidth,angle=270]{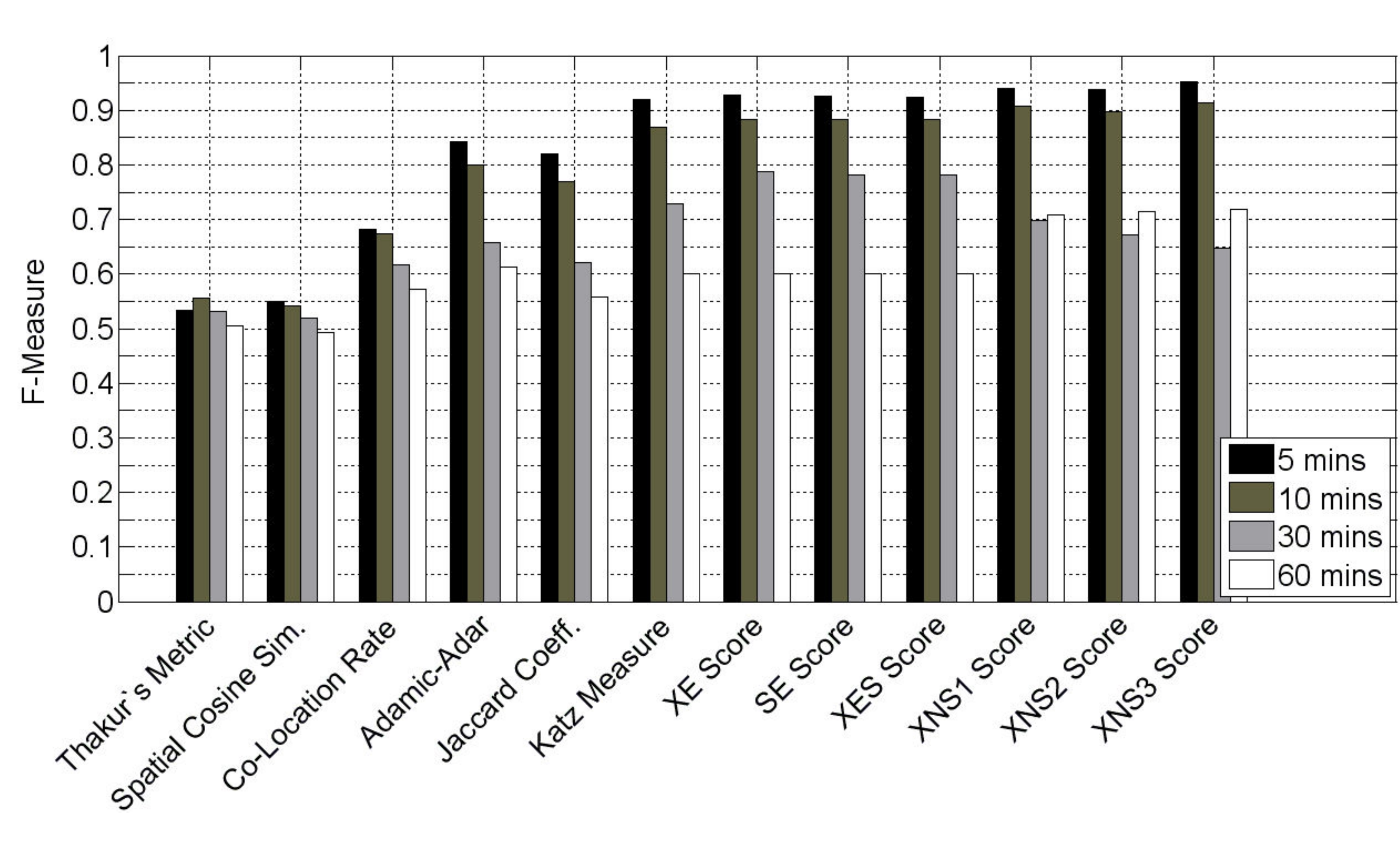}
  }
  \caption{Evaluation metrics for the prediction applied on the MIT Campus trace for different tracking periods}
  \label{Eval_Entropy_MIT}
\end{figure}

\begin{figure}[!t]
  \centering
  \subfigure[Top Scores Ratio in $T$+1]{
    \label{fig1}
    \includegraphics[width=0.3\textwidth,angle=270]{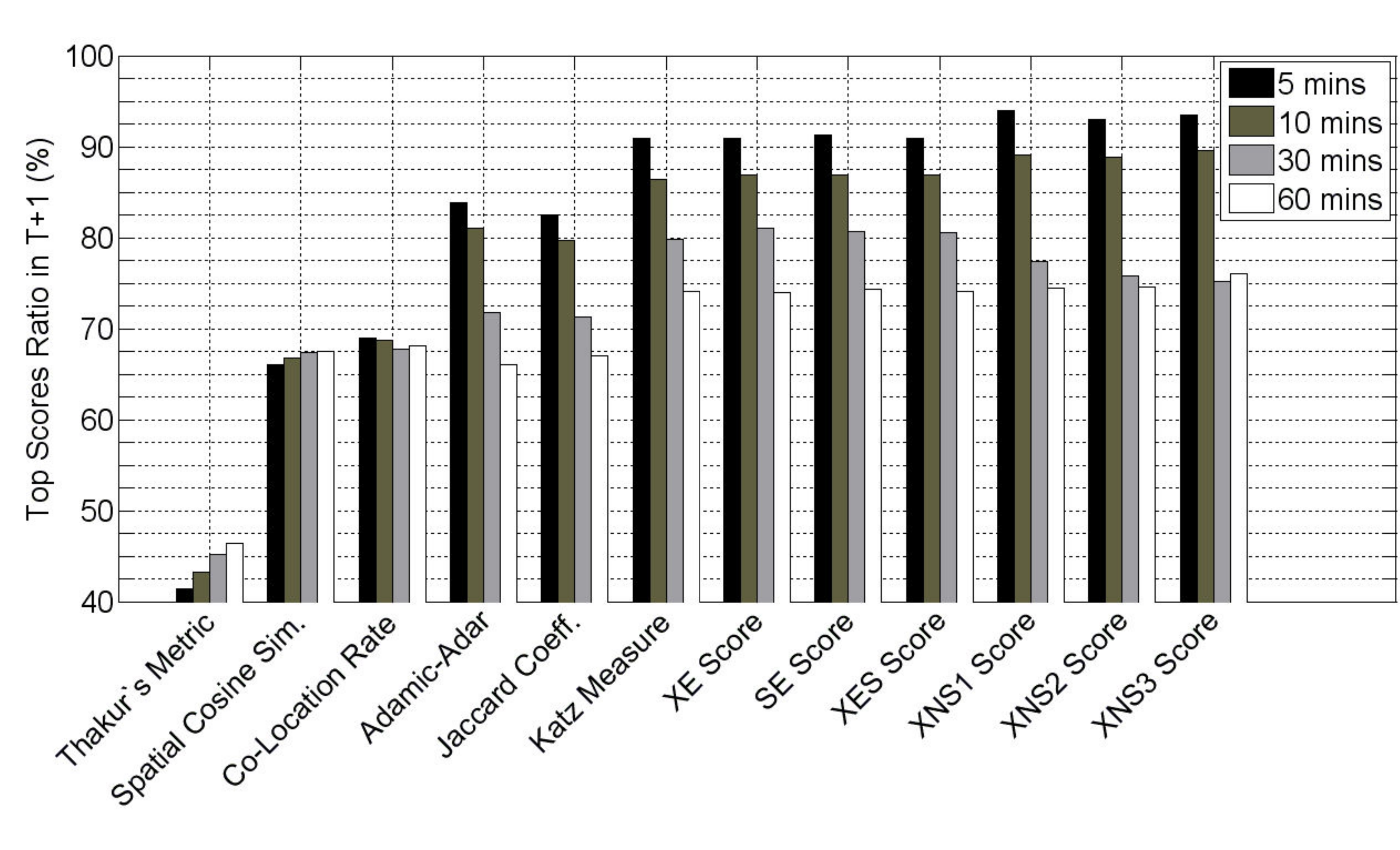}
  }\\
  \subfigure[Accuracy (percentage exceeding 99\%)]{
    \label{fig2}
    \includegraphics[width=0.3\textwidth,angle=270]{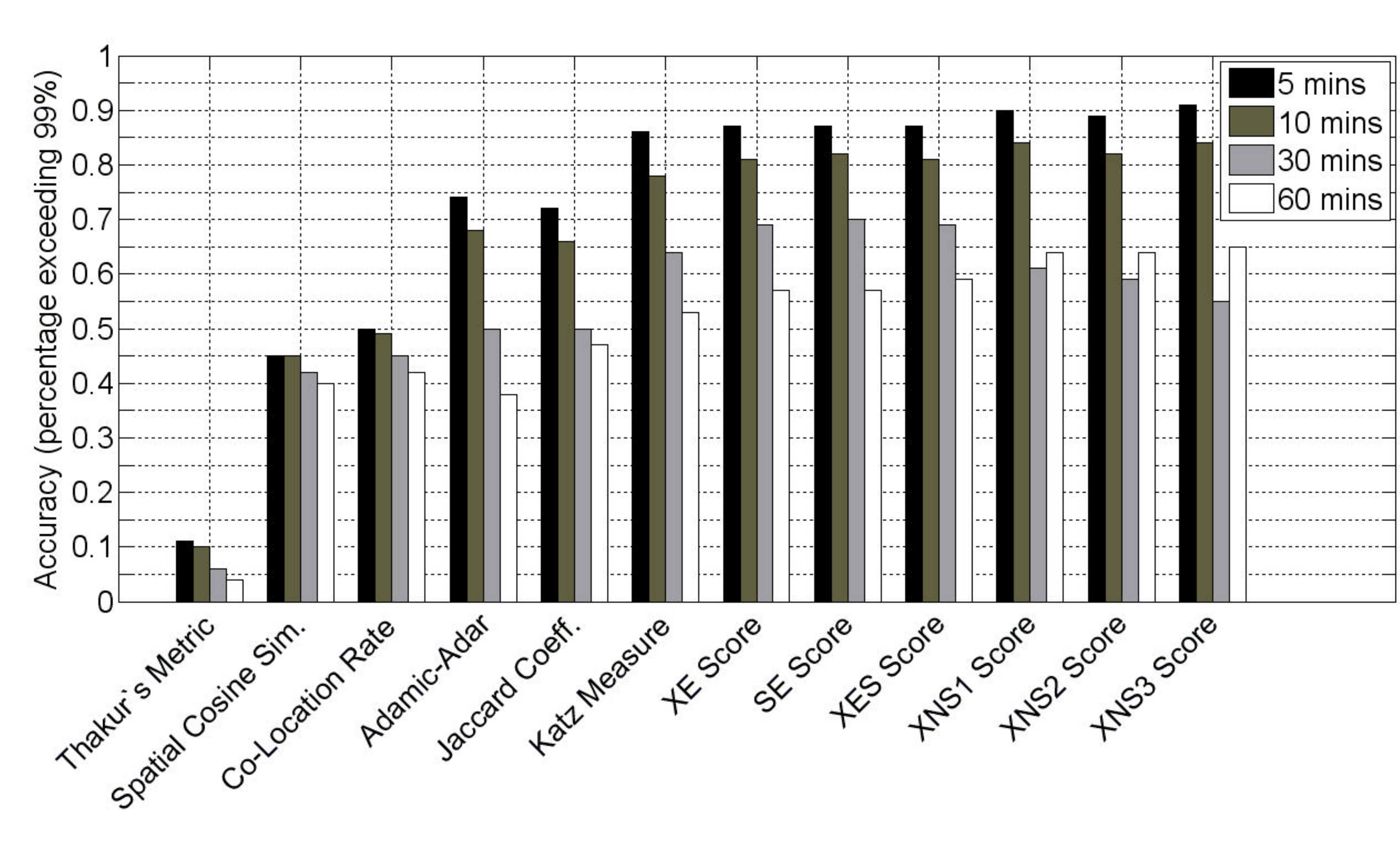}
  }\\
  \subfigure[F-Measure]{
    \label{fig3}
    \includegraphics[width=0.3\textwidth,angle=270]{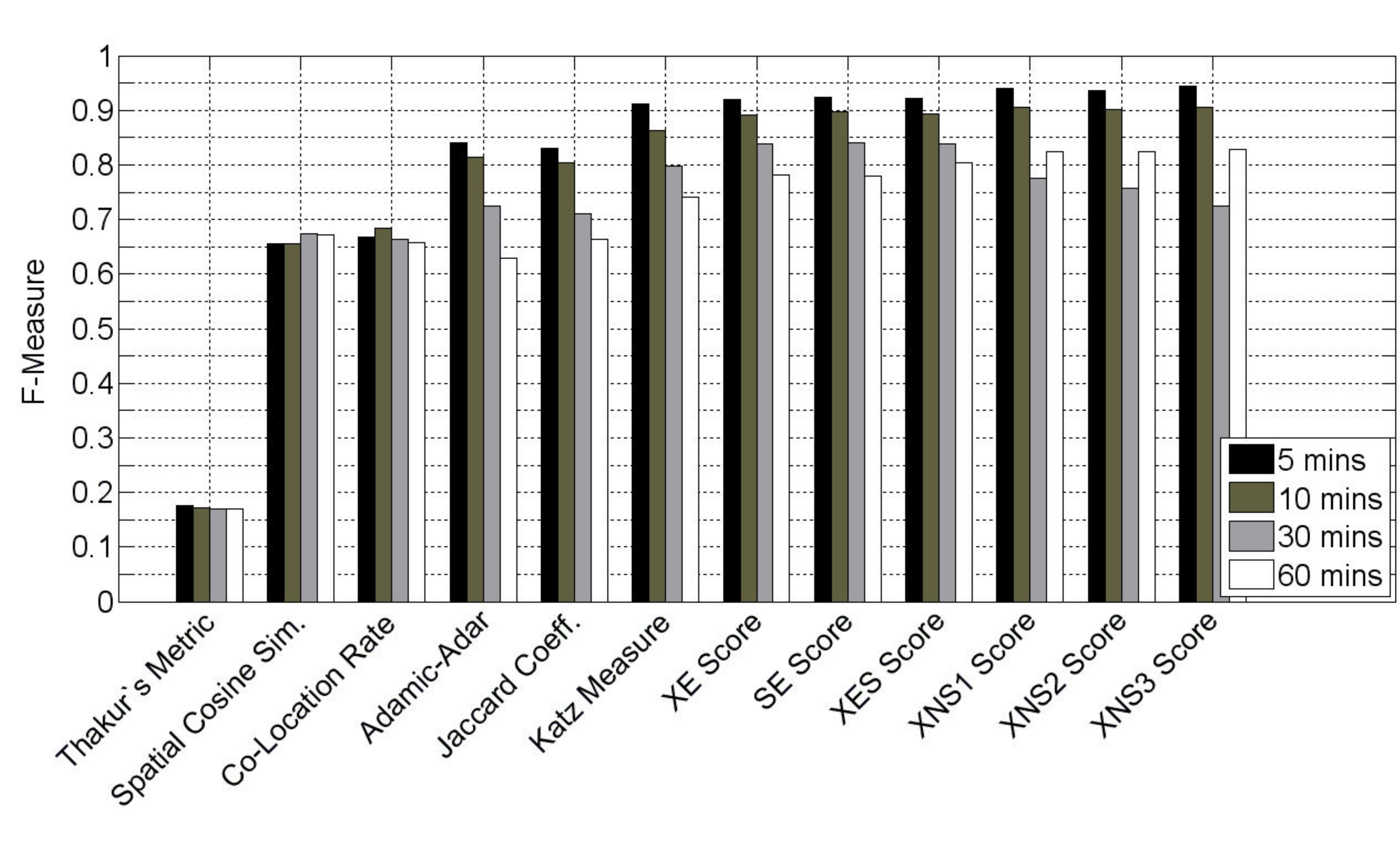}
  }
  \caption{Evaluation metrics for the prediction applied on the Dartmouth Campus trace for different tracking periods}
  \label{Eval_Entropy_Dartmouth}
\end{figure}

We compare the prediction efficiency of the proposed prediction
metrics with the one of the tensor-based link prediction technique
\cite{Zayani2011} based on the Katz measure. To propose a
comprehensive comparison, we also consider well-known prediction
metrics presented in the literature. On the one hand, we consider
behavioral-based link prediction metrics such as the similarity
metric of Thakur et al. \cite{Thakur2010} and two metrics expressing
mobile homophily proposed by Wang et al. in \cite{Wang2011}: the
spatial cosine similarity and the co-location rate. On the other
hand, we take two link prediction metrics based on measuring the
degree of proximity such as the Katz measure. They are the
Adamic-Adar measure \cite{Adamic2003} and the Jaccard's coefficient
\cite{Jaccard1901,Salton1986}. All the metrics that we propose are
involved in the latter category.

The evaluation metrics are computed for all traces with different
tracking periods lengths. Regarding the MIT Campus trace, the
results are reported in Fig. \ref{Eval_Entropy_MIT}. For the
Dartmouth Campus trace, the prediction results are listed in Fig.
\ref{Eval_Entropy_Dartmouth}. For each figure, the plots (a), (b)
and (c) respectively represent the top scores ratio in $T$+1, the
accuracy (the percentage exceeding 99\%) and the F-Measure obtained
for each prediction technique with different tracking periods.


The results obtained enable us to attest that the use of the Katz
measure has been one of the best choices to perform prediction
through the tensor-based technique. Using this metric achieves
better performance than those of the other link prediction metrics
proposed in the literature. Hence, the Katz measure is the best
metric that we can use to interpret the meaning of the entropy
estimations. The other well-known prediction metrics show a lower
performance in the context of predicting such links. Indeed, instead
of the Katz measure, they quantify the relationship of a pair of
nodes without seeing if a direct link connects them. For example,
two isolated neighbors (or having few common neighbors) would have a
weak score even if they were connected. On the other hand, two other
nodes can be 2-hop neighbors and share several common neighbors
which means that the score in this case is relatively high even if
no link connects the corresponding pair of nodes.

Comparing the performance of the tensor-based link prediction
technique with those of the proposed metrics based on link and/or
proximity entropy estimations leads us to assess that our proposal
of combining our prediction technique with stability measures is
coherent and effective. We remark that for each scenario, there is
at least three among the six proposed metrics that achieve better
prediction performance (better top scores ratio at $T+1$, accuracy
and F-measure) than the Katz measure. The Katz measure quantifies
the social ties between two nodes by a score. Nevertheless, such a
score sometimes cannot indicate if the contacts between these two
nodes are stable or interspersed over time. The contribution of the
entropy estimation is to identify the stable and persistent links
but not only them. Also, the entropy allocates low values when a
link never or rarely occurs between two nodes (long sequences of 0
for the status of a link). Hence, combining the Katz measure and the
entropy estimation enables us to distinguish more clearly a stable
link (low entropy estimation and high Katz measure). Such a
combination highlights more clearly how nodes are tied which makes
the prediction of the topology more efficient in the period $T$+1.

Considering both traces, we find that it is better to opt for the
$SE$, the $XES$ and the $XNS2$ scores when the tracking periods are
long (30 and 60 minutes). When these periods are shorter (10 and 5
minutes), the $XNS$ scores (particularly the $XNS3\_Score$) are the
most suitable metrics. In fact, shortening the length of the tensor
period leads to obtaining a more precise tracking of the properties
of the links (i.e. better distinction between persistent and
fleeting links), more faithful estimation of the stability and then
more efficient link prediction (more information about the impact of
the tracking period length is provided in \cite{Zayani2011}).
Meanwhile, when we use long tensor periods, tracking become less
precise as the method considers long and short contact as the same.
For example, if we consider a tracking period of 30 minutes, a
contact that occurs during all this period is considered as the same
as a contact that only lasts a few seconds during the same period
(the status of the link is set to 1). It is clear that we lose
precision when the contacts tend to be short. Moreover, we divide
the historical data into $T$ periods and when the tracking period
are longer, the number of total periods $T$ becomes less. Therefore,
the sequence of 0 and 1 that characterizes a link over time becomes
shorter and tends to be unstable (due to the lack of precision).
Afterwards, the Katz measure, as well as the entropy estimation, is
less efficient to characterize the relationship between nodes,
especially for short contacts.

\begin{figure}[!t]
  \centering

    \includegraphics[width=0.3\textwidth,angle=270]{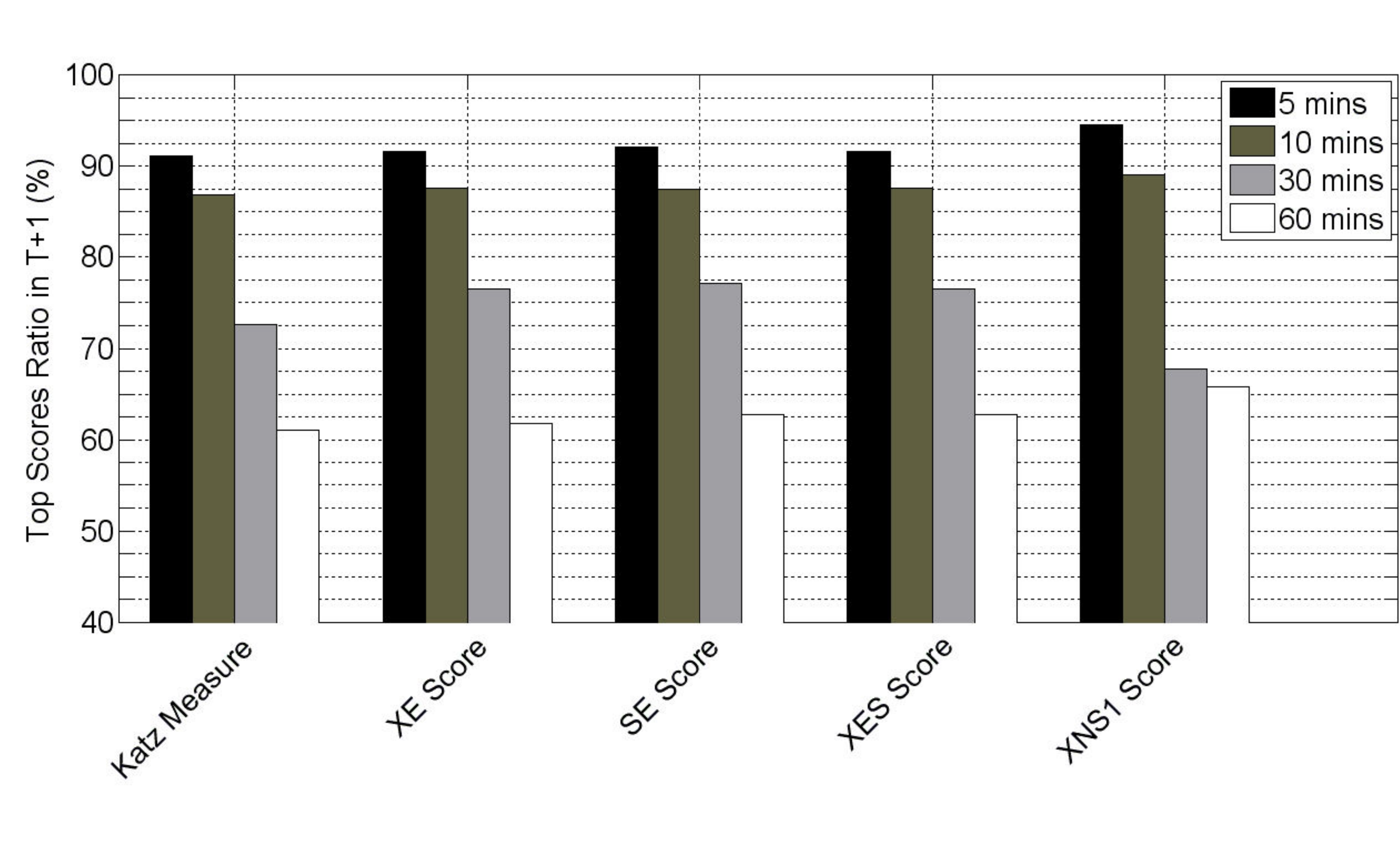}

  \caption{Top Scores Ratio in $T$+1 for the prediction applied on the MIT Campus trace with knowledge limited to direct neighbors and for different tracking periods}
  \label{Eval_Entropy_MIT_1hop}
\end{figure}

In these simulations, we have assumed that each node has knowledge
about its one-hop and two-hop neighbors. Limiting the local
information to one hop is not compromising the performance either of
the tensor-based link prediction technique or of the proposed
metrics based on entropy estimation. It is true that we are not
able, in this case, to compute the $XNS2$ and the $XNS3$ scores.
Nevertheless, the Katz measure is no longer affected with this
limitation. In fact, we have proved in \cite{Zayani2011} that the
prediction efficiency of the tensor-based link prediction technique
achieves similar performance in predicting links in the period $T$+1
whether the knowledge is limited to one-hop neighbors or extended to
the two-hop neighbors. Fig. \ref{Eval_Entropy_MIT_1hop} depicts the
Top Scores Ratio in $T$+1 obtained from the MIT Campus trace when
prediction is performed by the Katz measure and by the proposed link
stability based measures using only the knowledge of direct
neighbors ($XE$, $SE$, $XES$ and $XNS1$ scores).
The results confirm the findings cited above: the performance of
each prediction technique is similar in both scenarios. Obviously,
the contribution of the entropy estimation is always effective.

\section{Conclusion}

In this paper, we highlighted that the prediction efficiency of the
tensor-based technique that we have advanced can be improved by
taking into consideration other aspects on top of measuring social
closeness. We proposed to make the prediction sensitive to the link
and proximity stabilities. We showed that a strong relationship
between two nodes matches also with a stable link between them.
Indeed, when similar intentions are shared for a period of time, the
link between the corresponding individuals is expected to be stable.
We also outlined that considering proximity stability can also be
beneficial to improving the prediction performance. To express the
stability itself, we proposed an entropy estimator inspired from
data compression and which converges to the expected value of
entropy for a time series. In our case, the time series is the
sequence of the state of the links in the adjacency matrices
(sequence of zeros and/or ones).

To assess the efficiency of our contribution, we tried to join the
stability feedback to the tensor-based link prediction framework
through proposing new prediction metrics. We assessed that we can
improve the performance of the prediction technique especially when
the tensor time period tend to be short. In other words, using
shorter tensor periods favors more precise tracking of contacts
between nodes, which leads to a better estimation of the entropy
estimations and then more faithful link prediction. Above all, we
identified, according to the length of tracking period, the set of
metrics that can be used to enhance, as much as possible, the
performance of the tensor-based link prediction technique.

As future perspectives, it will be interesting to focus on the best
combination that we can design in order to reach the highest level
of prediction efficiency. In this work, we have only proposed some
metrics to highlight that measuring the stability of ties can be
beneficial without seeking the best combination in terms of
prediction performance. In addition, it would be challenging to join
our contribution with some communication protocols and especially
with opportunistic routing protocols. We aspire to verify that the
feedback provided by our contribution will improve the performance
of such protocols through guiding them in taking better decisions by
using link prediction, to forward packets.

\section*{Acknowledgements}
We want to thank wholeheartedly Makhlouf Hadji for his helpful
advice.

\bibliographystyle{ieeetran}
\bibliography{Biblio}

\end{document}